\pdfoutput=1
\documentclass[12pt]{article}
\usepackage{graphicx}
\usepackage{geometry} % see geometry.pdf on how to lay out the page. There's lots.
\title{The Planck-LFI flight model composite waveguides}
\author{O. D'Arcangelo$^1$, L. Figini$^1$, A. Simonetto$^1$, F.Villa$^2$, \\M. Pecora$^3$, P. Battaglia$^3$, M. Bersanelli$^4$, R.~C. Butler$^2$,\\ F. Cuttaia$^2$, S. Garavaglia$^1$, P. Guzzi$^3$, N. Mandolesi$^2$,\\ A. Mennella$^4$, G. Morgante$^2$, L. Pagan$^3$ and L. Valenziano$^2$  }
\date{} % delete this line to display the current date
\begin{document}

\maketitle
%\tableofcontents abstract
\begin{center}
$^1$\emph{Istituto di Fisica del Plasma - CNR, via Cozzi 53, 20125 Milano, Italy} \\ 
$^2$\emph{Istituto di Astrofisica Spaziale e Fisica Cosmica, INAF, via P. Gobetti, 101, I40129 Bologna, Italy}\\
$^3$\emph{Thales Alenia Space Italia, s.s. Padana Superiore 290, 20090 Vimodrone (MI), Italy}\\
$^4$\emph{Universit\`a degli Studi di Milano, Via Celoria 16, 20133 Milano, Italy}
\end{center}
\begin{center} \textbf{Abstract}
\end{center}
The Low Frequency Instrument on board the PLANCK satellite is designed to give  the most accurate map ever of the CMB anisotropy of the whole sky over a broad frequency band spanning 27 to 77 GHz. It is  made of an array of 22 pseudo-correlation radiometers, composed of 11 actively cooled (20 K) Front End Modules (FEMs), and 11 Back End Modules (BEMs) at 300K, each FEM and BEM set comprising two radiometers. The connection between the two  parts is  made with rectangular Wave Guides (WGs).  
Considerations of very different nature (thermal, electromagnetic and mechanical), imposed stringent requirements on the WGs characteristics and drove their design. 
From the thermal point of view, the WG should guarantee good insulation between the FEM and the BEM sections to avoid overloading the cryocooler. 
On the other hand it is essential that the signals do not undergo excessive attenuation through the WG. 
Finally, given the different positions of the FEM modules behind the focal surface and the mechanical constraints given by the surrounding structures, different mechanical designs  were necessary. A composite configuration of Stainless Steel and Copper  was selected to satisfy all the  requirements described. Given the complex shape and the considerable length (about 1.5-2 m) of the LFI WGs, manufacturing and testing the WGs was a challenge. This work deals with the development of the LFI WGs, including  the choice of the final configuration and of the fabrication process.  It also describes the testing procedure adopted to fully characterize these components from the electromagnetic point of view and the space qualification process they underwent. The Scattering parameters of the WGs were obtained in a one port configuration, minimizing instrumental errors. The space qualification process required manufacturing ad-hoc facilities to support the WGs during vibration tests and to allow electromagnetic tests without removing them. Results obtained during the test campaign are reported and compared with the stringent requirements. The performance of the LFI WGs is in line with requirements, and the WGs were successfully space qualified.

\section{Introduction}
The PLANCK satellite, ESA's third generation space mission devoted to 
the study of the Cosmic Microwave Background (CMB), is designed to produce 
a map of the CMB
anisotropy over the whole sky, with an unprecedented combination of
angular resolution (4'-30') and sensitivity 
($\Delta$T/T$\simeq10^{-6}$), for a wide range of frequencies
(27-850 GHz). Two complementary instruments, LFI ($Low$
$Frequency$ $Instrument$) operating in 3 20$\%$  wide frequency bands, 
centered at 30, 44 and 70GHz (Bersanelli et al. \cite{bersanelli})
and HFI ($High$ $Frequency$ $Instrument$ in the 100-850 GHz 
range) (Puget et al. \cite{puget}), have been recently integrated together and  
share the focal plane of a Gregorian off-axis aplanatic optimized 
telescope (Villa et al. 2002).

LFI consists of an array of 22 wide-band pseudo-correlation 
radiometers (Cuttaia et al. \cite{cuttaia}), composed of a Front End Unit (FEU) working at 20 K (Davis et al. \cite{davis}; Varis et al.\cite{varis}),  where 
the signal is collected and amplified, and a Back End Unit (BEU) (Artal et al. \cite{artal}; Varis et al. \cite{varis}), working 
at  300K, where the signal is further amplified and then detected. The LFI 
WGs are necessary in order to connect the two parts of the 
radiometers. 

\begin{figure}[!h]
\begin{center}
\includegraphics[width=0.6\textwidth]{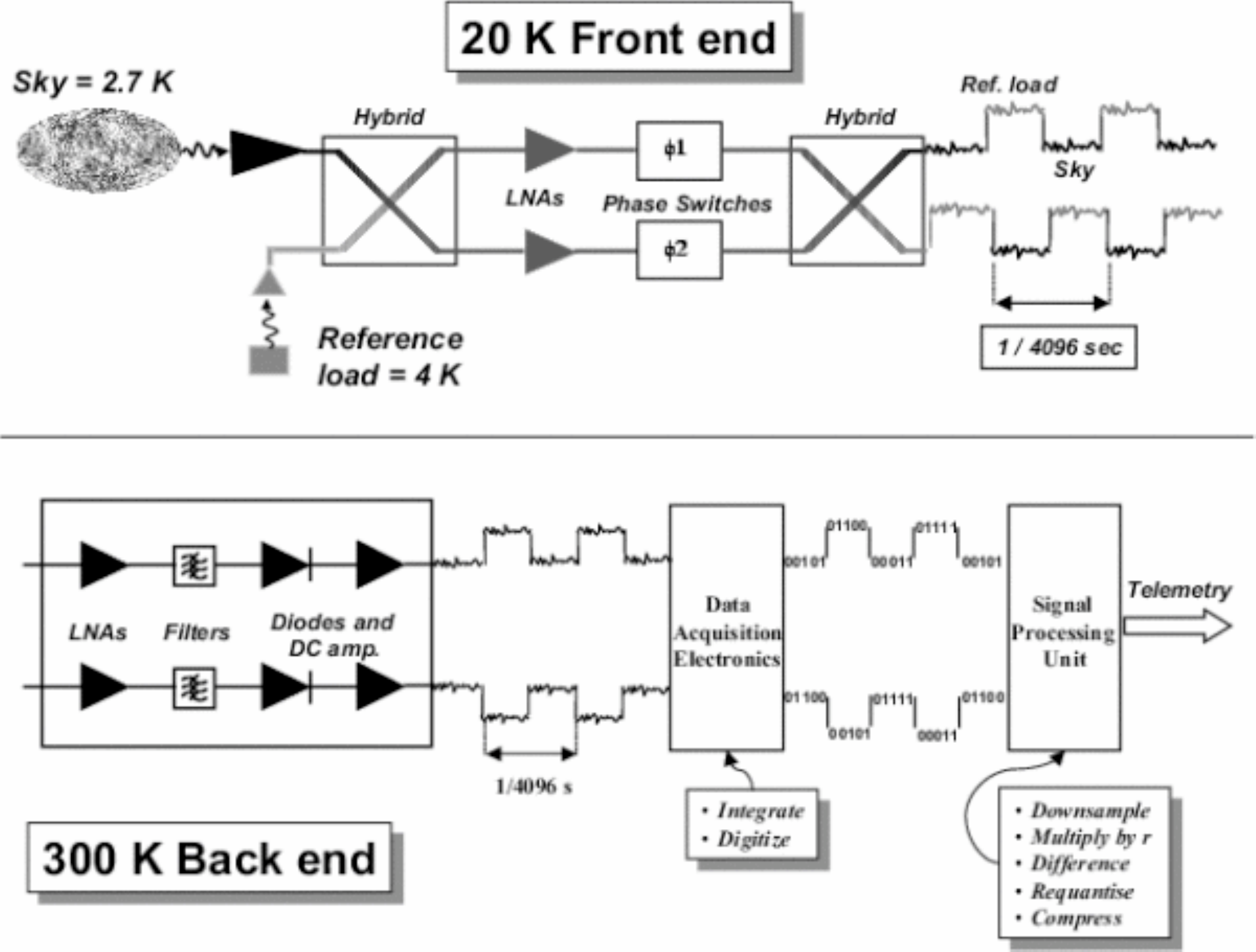}
\caption{\emph{Scheme of the PLANCK-LFI radiometers.}}
\end{center}
\label{fig:radiometer_scheme}
\end{figure}

The FEU is composed of 11 dual profile corrugated  Feed Horns (FHs) (Villa et al. \cite{villa}), placed 
in the outermost region of the PLANCK focal plane, each one connected to 
an Ortho Mode Transducer (OMT) (D'Arcangelo et al. \cite{D'Arcangelo}) that splits the incoming radiation in two 
orthogonal, linearly polarized signals, each feeding two independent 
Radiometric Chain Assemblies (RCA). A Front End Module (FEM) 
is mounted behind each OMT: here the signal coming from the sky is continuously 
compared with a reference black body signal at 4 K provided by a pyramidal 
horn facing a microwave load (Valenziano et al. \cite{valenziano}). Every FEM has 4 outputs, two for each RCA.  
Thus the 44 LFI WGs, arranged in groups of 4, connect the 20 K FEU and the 300 K 
BEU, covering a distance between about 1.5 and 2 m. A schematic view of 
an LFI radiometer is given in figure \ref{fig:radiometer_scheme}. The goal of this paper is to report the design, fabrication and testing activity of the WG: the results obtained during the tests are thus compared with the specification requirements.
The paper is organized as follows: in sections 2 and 3 the design and fabrication technique of the WGs are described.
In section 4 the electromagnetic tests made on the WGs and the measurement 
techniques are presented and discussed. 
In section 5 the results of the Flight Model (FM) test campaign are shown, while 
section 6 deals with vibration tests.
Section 7 provides a summary and a few concluding remarks.

\section{Design}
\begin{figure}[!h]
\begin{center}
\includegraphics[width=0.6\textwidth]{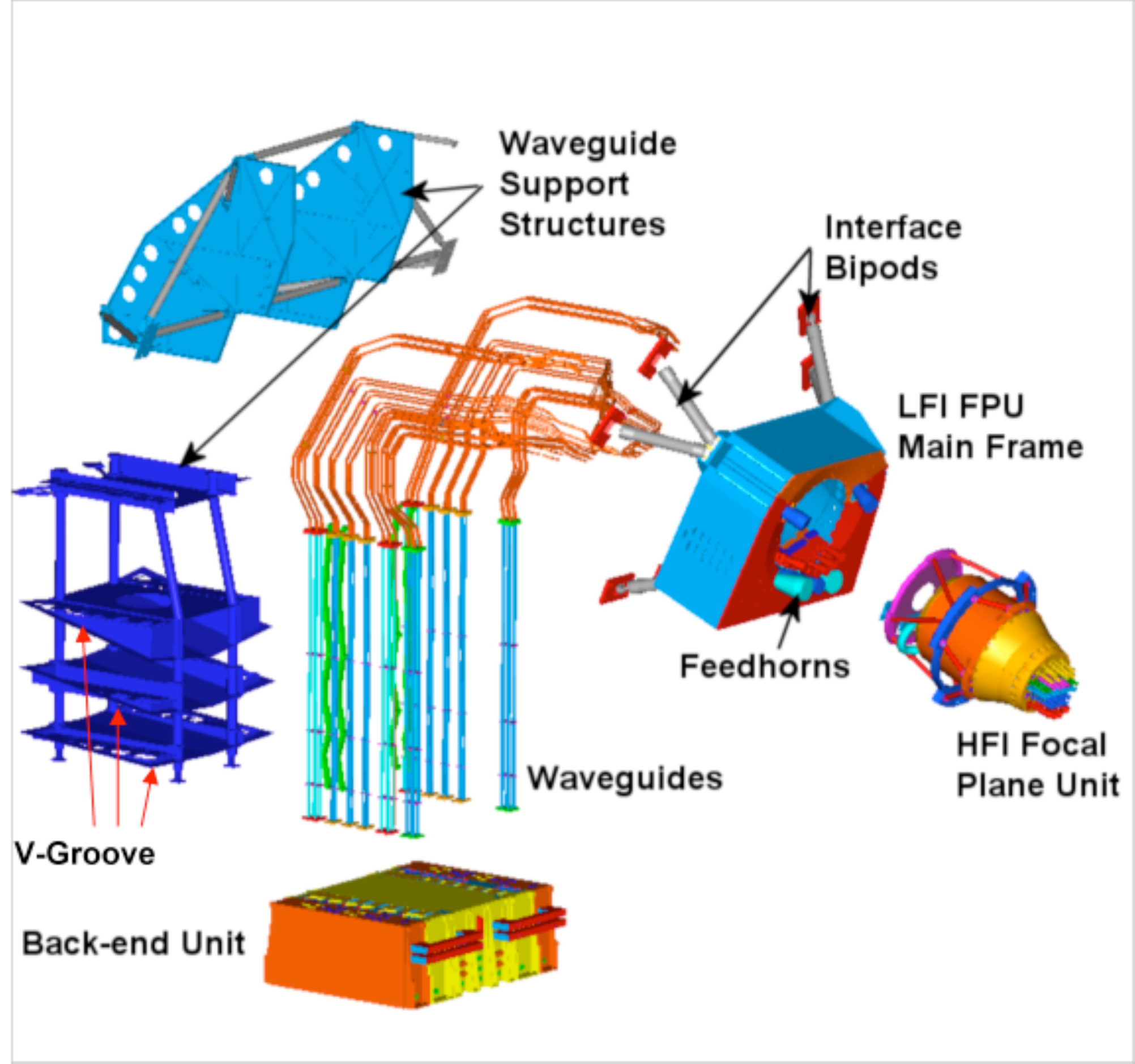}
\caption{\emph{Schematic view of the PLANCK LFI/HFI}} 
\end{center}
\label{fig:LFI}
\end{figure}

\begin{figure}[!h]
\begin{center}
\includegraphics[width=0.8\textwidth]{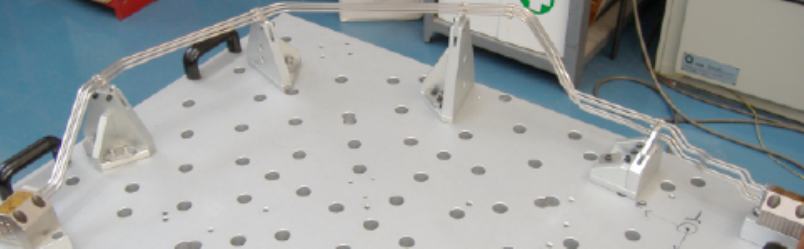}
\caption{\emph{Copper 70 GHz WGs}} 
\label{fig:Cu70GHz}
\end{center}
\end{figure}

Metallic waveguides are commonly used in microwave technologies as transmission lines. 
They consist of metal hollow pipes usually with rectangular cross section,guiding electromagnetic waves. 
The most commonly used are rectangular in section with standard dimensions. For an exhaustive treatment of waveguides see 
Marcuvitz (\cite{marcuvitz}) and Cronin (\cite{cronin}). For LFI it was decided to use standard rectangular waveguides to reduce costs and manufacturing criticalities. 
Their unimodal propagation simplifies matching with the MMIC (Millimeter Wave Integrated Circuits) of the FEM and control of discontinuities. Losses are higher than in oversized waveguide but they are still within the performance budget (Bersanelli, \cite{bersanelli}).
The complexity of LFI translated into the waveguides that required extensive prototyping activity and development strategy for design, manufacturing, measurements and space qualification.
As usual for space components, LFI WGs must exhibit low mass and 
compatibility with payload vibration at launch, but their characteristics 
are also strongly driven by the role they play in the radiometer structure, 
since they interconnect two parts of the 
instrument operating at very different temperatures. Thus it is essential 
for them to have a very low thermal conductivity in order to decouple the FEM 
at 20 K and the BEM at 300 K. On the other hand, they also must show good 
electromagnetic performance, i.e. acceptable Insertion Loss (IL) and high 
Return Loss (RL), in order to allow the radiometric signal to reach the 
last stage of detection without excessive attenuation. Finally, they 
require a complex geometric routing, since each of them must reach a 
different location in the focal surface, while allowing 
integration of HFI in the LFI frame and of both instruments 
on the satellite (figure \ref{fig:LFI}). 
Thus the final configuration chosen for the WGs design is a tradeoff 
between thermal, electromagnetic and mechanical aspects; the first 
consequence is that the WGs were built using 
different materials. They are composed of two separate sections (connected by non standard flanges), one made 
of Stainless Steel (SS) and the other made of Copper (Cu), both of 
standard rectangular size. 
SS sections have the same length (700mm)  at all frequencies, and a very strong thermal gradient along their path. The inner part is partially gold plated to minimize losses. The Cu section is the one that reaches the FEU, and it accommodates all the 
bends and twists necessary to reach different positions of the FEU and to take 
into account the surrounding structures that impose complex geometric 
routing. They will work at a constant temperature of 20 K. Copper was selected because of its  malleability that 
minimizes the risk of mechanical damage. It has also a good electrical 
conductivity at cryogenic temperature and can be easily electroformed. The length and shape of these WG sections is variable  between about 800 and 1200 mm, depending on the different positions reached 
in the FEU. Figure \ref{fig:Cu70GHz} shows 4 Cu WGs at 70 GHz . 
At the beginning of the waveguide development, it was foreseen to avoid any interconnection flange between the different waveguide sections as in the case of WMAP (Jarosik et al., \cite{n.jarosik}). However, the extremely complex waveguide path and the 2 m length did not allow building the waveguide in a single piece.

\subsection{Electromagnetic design}
The copper section bends and twists were designed optimizing the return loss using analytical formulae in Johnson, \cite{r.c.johnson}. Then the analytical approach was verified with plane finite element software (HFFS\footnote{http://www.ansoft.com}) for the typical case of H-- and E-- bends both isolated and cascaded. 
For E--plane bends the analytic form used for the reflection coefficient is 
\begin{equation}
\left | {\rho}\right | = {b^2\over 24 R^2} \left | \sin \left ( {2\pi \ell \over \lambda _g }\right ) - A\cdot 
\cos \left ({2\pi \ell \over \lambda _g } \right) \right |
, \qquad \mathrm{for}\, {R\over b} > 2
\end{equation}
where $R$ is the radius of curvature of the bend, $\ell$ is the length of the bend along its path, $\lambda_g$ is the wavelength inside the waveguide, $b$ is the short side size of the waveguide. The parameter $A$ is calculated as follows: 

\begin{equation}
A = \left( {96 \over \pi^4} \right)\left({2 b \over \lambda_g}\right) \sum^{\infty}_{1,3} m^{-5} \left[ \-\left({2b\over m\lambda_g}\right)^2\right]^{-1/2} 
\end{equation}

The best match is obtained if the length of the bend satisfies the following condition: 
\begin{equation}
\ell = R \cdot \theta = {\lambda_g \over 2\pi} \tan^{-1} A
\end{equation}
In addition the length $\ell$ can be increased in steps of $\lambda_g/2$ in order to arrive at an allowed value of $R/b$. Although this is a resonant condition, the simulations and the measurements on all the LFI waveguides showed that these analytical formulas can be used to design complex waveguides. 
Similarly for H--plane bends the formulas become

\begin{equation}
\left | {\rho}\right | = {\lambda_g^2\over 32\pi^2 R^2} \left | \sin \left ( {2\pi \ell \over \lambda _g }\right ) - {128 \over \pi^2} {a\over\lambda_g} B \cdot 
\cos \left ({2\pi \ell \over \lambda _g } \right) \right |
, \quad \textrm{for}\,{R\over a} > 2
\end{equation}
\begin{equation}
B = \sum^{\infty}_{2,4} m^2 \left(m^2-1\right)^{-3} \left[{\left(m^2-1\right)}-\left({2a\over\lambda_g}\right)^2\right]^{-1/2}
\end{equation}
\begin{equation}
\ell = R \cdot \theta = {\lambda_g \over 2\pi} \tan^{-1} \left[\left({128 \over \pi^2} \right) {a\over \lambda_g} B\right]
\end{equation} 
where $a$ is the long side size of the waveguide.

The length of the twist was chosen so that

\begin{equation}
\ell = \left ( 2 n \right ) 0.25 \lambda_t 
\end{equation}
where $n$ is an integer and $\lambda_t$ is the wavelength inside the twist. 

To improve the return loss of the copper sections, a straight section at least one wavelength long has been interposed between subsequent discontinuities. This was done in order to reestablish the $TE_{10}$ fundamental propagation mode in case of unwanted spurious mode excitation and prevent tunneling of higher order modes through discontinuities. 
During the design phase it was also established to insert a maximum of one twist for each waveguide and to use a long radius of curvature where possible, again with the goal of improving return loss.
Manufacturing tolerances were studied based on Alison, (\cite{w.b.w.alison}) and set after a prototype activity on WR10 (W--band) waveguides. This included also the verification of the electromagnetic properties and methods to guarantee a return loss better than $30$dB over the entire bandwidth and an attenuation as good as straight waveguides. 
 
\subsection{Thermal design}
The thermal characteristics of the WGs played an important role in the final design configuration: in fact, since the LFI waveguides connect the warm, $\sim$300~K unit to the cold front end of the instrument, along their routing they have been connected mechanically and thermally to the three thermal shields (V-grooves) of Planck, passively thermalised at $\sim$150~K, $\sim$100~K and $\sim$50~K. Therefore heat transfer and electrical conduction played a key 
role in the waveguide design, with the goal of keeping the heat load to the 20~K stage
at the level of $\sim$100~mW and the insertion loss at a level of a few dBs. The composite design described in detail in this section has been therefore
devised based on an analytical heat transfer model
\footnote{Radiative heat transfer inside waveguides is negligible compared to conduction because (i)
the inner waveguide surface is far from being specular to IR radiation and (ii) twists and bends 
favour absorption of IR photons}. 
To estimate the heat transferred by conduction and, therefore, the temperature profile along 
the waveguide, we have formalised the problem by making simplifying assumptions, i.e.:

\begin{itemize}
   \item the thermal link between waveguides and thermal shields is assumed to be perfect, 
   which implies that all the heat coming from higher temperature regions will be transferred 
   to the V-groove;
   \item every V-groove can be considered as an infinite thermal sink;
   \item no heat will be dissipated by the waveguides between two V-grooves.
\end{itemize}

\begin{figure}[h!]
   \begin{center}
      \includegraphics[width=7cm]{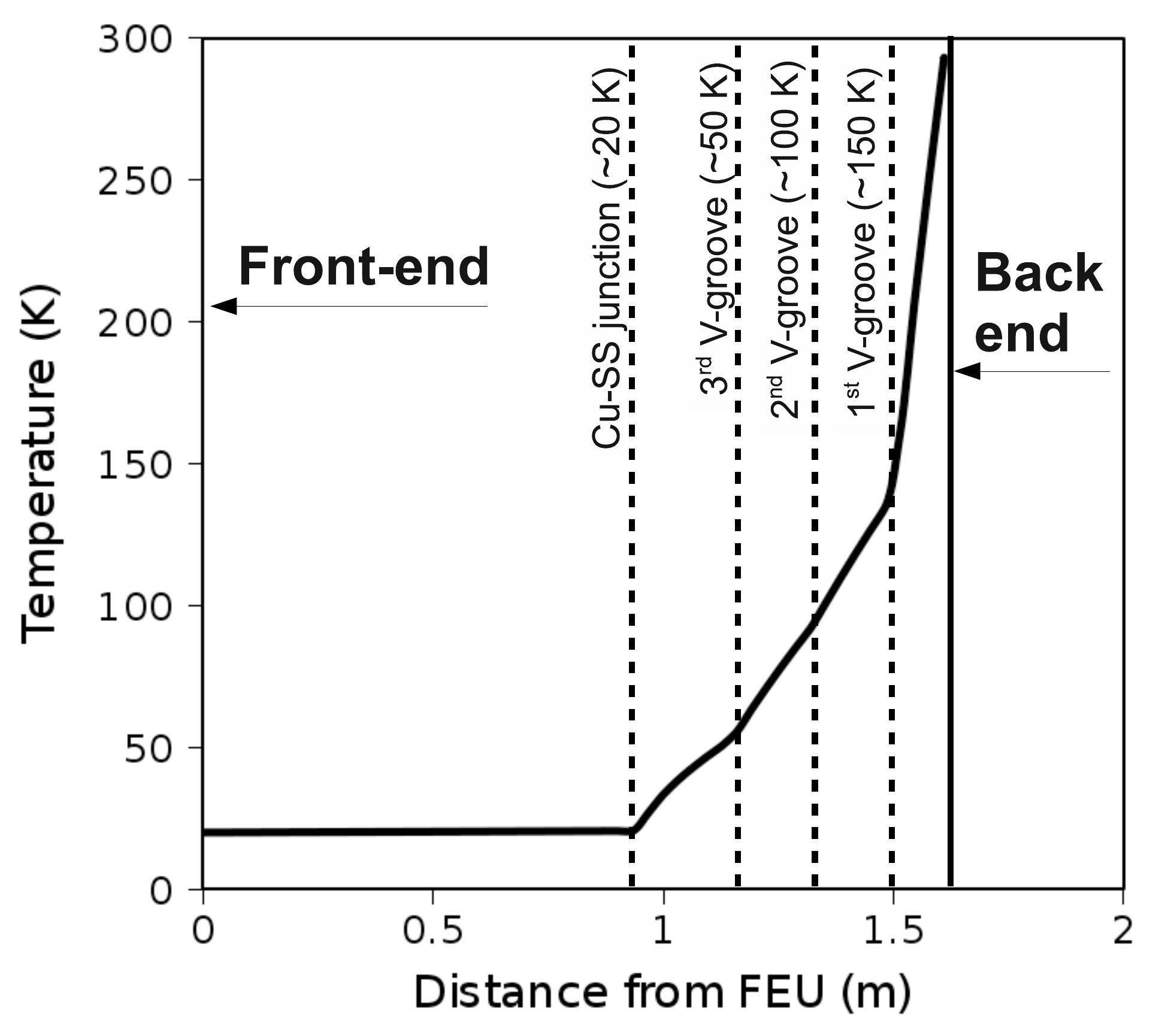}
   \end{center}
   \caption{Temperature along the LFI waveguides calculated with 
   Eq.~\ref{eq:qdot_cond}}
   \label{fig:temperature_profile}
\end{figure}

Under these assumptions the heat load ($\dot Q$) on a given thermal link set at temperature 
$T_{\rm low}$ is constant, and determined uniquely by the waveguide characteristics and by the 
temperatures at the extremes ($T_{\rm low}$, $T_{\rm high}$), i.e.:
\begin{equation}
    \dot Q^{\rm cond}= \frac{A}{l}  \int_{T_{\rm low}}^{T_{\rm high}} k(T){\rm d} T
    \label{eq:qdot_cond}
\end{equation}

where $A$ is the waveguide wall section, $l$ represents the distance between thermal interfaces at 
$T_{\rm low}$ and $T_{\rm high}$ and $k(T)$ is the thermal conductivity function. 
From equation \ref{eq:qdot_cond} it is possible to derive the temperature profile along the
waveguide, considering that at any point $\dot Q^\mathrm{cond}$ is constant and, therefore
\begin{equation}
\dot Q^{\rm cond}= \frac{A}{x}  \int_{T_{\rm low}}^{T(x)} k(T){\rm d} T
    \label{eq:qdot_cond1}
\end{equation}

for each value of $x$. From a computational point of view the easiest way to calculate $T(x)$ has been to generate a list of temperature values in the range $[T_{\rm low}, T_{\rm high}]$ and then 
calculate the corresponding values of $x$ from the above equation. In Fig~\ref{fig:temperature_profile} we show an example of temperature profile calculated 
using Eq.~\ref{eq:qdot_cond}.
The last question considered during the design phase was the necessity for safely reaching vacuum conditions in space, since LFI waveguides may create problem in evacuating the LFI frontÐend and backÐend modules during the satellite launch. In order to avoid damages to the radiometers, the WG flanges include venting holes to aid the flow of air. Holes in the waveguide wall behave as a third port, producing mainly two effects:  they create discontinuities (generating reflections inside the waveguide) and they radiate electromagnetic power into space.
Reflections can be limited by an appropriate location of the hole where perturbation to wall currents is minimum, while radiation can be controlled by keeping the hole diameter below cutoff for the fundamental circular waveguide, i.e.: 

\begin{equation}
   d<<d_{max}=\frac{1.841}{\pi\sqrt{\epsilon_{0}\mu_{0}}\nu_{hf}}
\end{equation}
where $\nu_{hf}$ represents the highest frequency propagating inside the WG, 77 GHz for the LFI instrument, which corresponds to $d_{max}$=2.3 mm.
Hence, holes dimension must come out as a trade-off between fluid conductance and electromagnetic requirements.
The resulting optimized drawing foresees a single hole per waveguide, directly on the flange, with the hole axis parallel to the HÐplane. The hole diameters is 1 mm: this value looks large enough for the venting purpose and safe from the electromagnetic point of view. The length of the hole is  6.95 mm. Simulations, performed at the highest in band frequency, demonstrated that the holes do not impact the electromagnetic performance of the WG and that the power radiated through the hole is well below the measurable level.

\section{Fabrication technique}
\begin{figure}[!h]
\begin{center}
\includegraphics[width=0.6\textwidth]{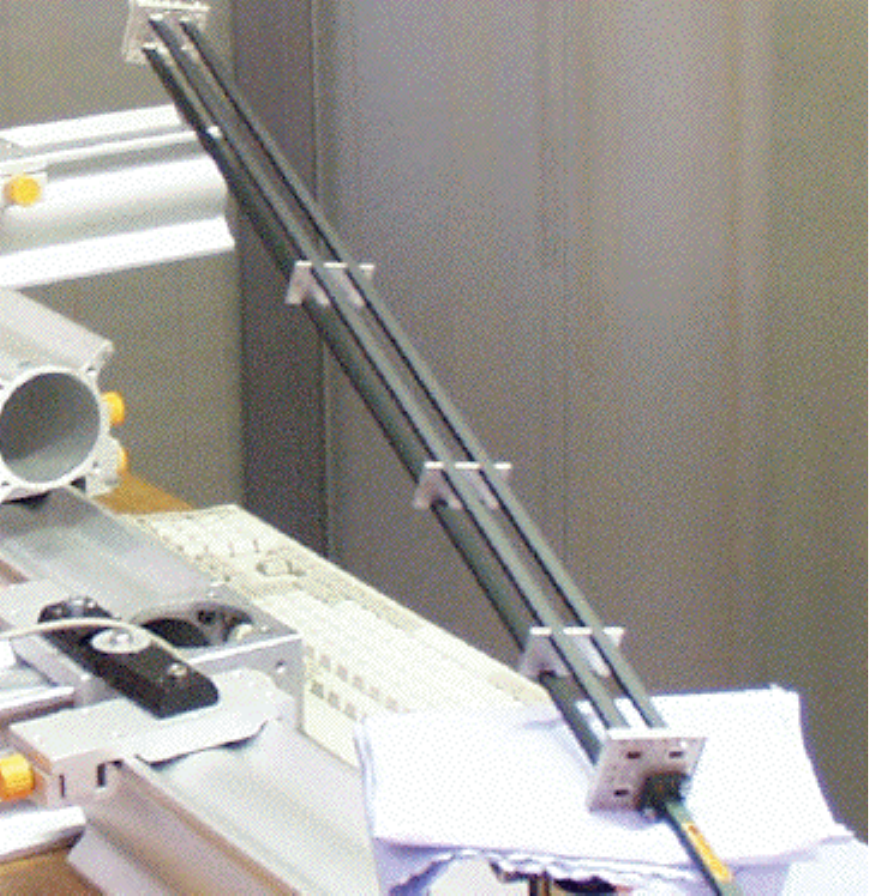}
\caption{\emph{Stainless Steel 30 GHz WG}} 
\label{fig:SS30GHz}
\end{center}
\end{figure}

\begin{figure}[!h]
\begin{center}
\includegraphics[width=0.6\textwidth]{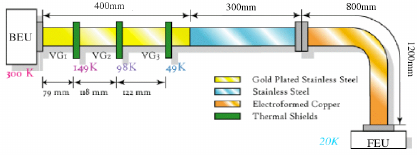}
\caption{\emph{Schematic view of the inner part of the LFI's WG}} 
\label{fig:schemaWG_color}
\end{center}
\end{figure}

The WG section between the Cu WG and the BEU is made, for the first 300 mm, of stainless steel, whose poor thermal conductivity limits the thermal input on the 20 K stage. Since SS is also a poor electrical conductor, the subsequent 400 mm of the WG are gold plated internally, in order 
to minimize signal losses, while preserving a low thermal conductivity. Thermal decoupling is a critical issue for LFI, so each WG is connected to three thermal shields (V-grooves) to enhance control of the heat load on the FEU.
In addition, a paint treatment with Aeroglaze Z306, a high emissivity (0.94-0.97) and low outgassing black polyurethane,  was used for the SS section. This paint provides the thermo-optical properties required to dissipate as much as possible into space. The emissivity of Aeroglaze Z306 increases with temperature, thus further preserving the coldest part of the satellite from radiative coupling.
 The SS sections are straight rectangular WGs with the same length (700 mm) at 
all frequencies. An image of four 30 GHz SS  WGs  is shown in figure \ref{fig:SS30GHz}. 
A schematic reproduction of the inner part of the two WG sections is shown 
in figure \ref{fig:schemaWG_color}, while the physical characteristics for the selected WG's design are reported in table \ref{Tab:1}.

\begin{table}[hb!]
       \centering
       \begin{tabular}{lccccc}
       \hline\hline

Central  &	Internal&        &   Thickness&\\
Frequency &section
&Cu   	&SS  &	Au \\
 $[GHz]$& [mm]  & [mm]& [mm]& [$\mu$m]\\
 30     & 7.112x3.556 	& 0.4 	&0.254&	2.0   \\
 44     &   5.690x2.840  &	 0.4 	&0.254 &	2.0 \\
 70      &   3.099x1.549 &	 0.4 	&0.254  &      2.0  \\
       \hline\hline
       \end{tabular}
       \caption{\emph{Mechanical properties of the LFI's WGs }}
       \label{Tab:1}
\end{table}
The two sections of the WGs are connected with a custom 
flange. The connection between the Cu section and the FEU requires a non 
standard flange at all frequencies, while the connection between the SS section and the BEU requires non standard flanges only at 44 and 70 GHz.
The two separate WG sections show different criticalities, since 
they are made of different materials and have different shape and length. 
The main difficulty in manufacuring the SS section consists in the 
partial gold plating of its inner part, especially at higher frequencies. 
For the Cu section, instead, the main problem is to make the bends and 
twists requested by the design. 
Two different manufacturing processes were selected for the SS and the Cu 
sections. The SS parts were directly built in a standard way and the gold 
deposition was made by placing the anode wire inside the WG, ensuring that it did not hit the internal walls, 
and providing the necessary 
flux of electrolyte inside the WG. The part that was not to be 
plated was masked before the deposition. 
The Cu section were instead built using the electroforming technique, even if it is a very complicated process. During the design phase, 
the contribution of discontinuities such as E and H plane 
bends and twists  to losses was estimated analytically. Calculations showed  that any possible distortion of the WG size has a larger impact on the RL of the component than the bends and twists themselves (LFI scientific team, 2003). Also the routing of these sections was considered and optimized during the design phase: the number of twists was minimized because of the criticality in manufacturing the mandrels, and the number of different bends in order to simplify mandrel integration before electroforming. Other criteria were the minimization of WG length and the use of straight sections wherever possible in order to minimize RF losses. 
The tolerances declared by the manufacturer are reported in tables  \ref{Tab:2} (Cu section) and  \ref{Tab:3} (SS section).
\begin{table}[hb!]
       \centering
       \begin{tabular}{lcccc}
       \hline\hline
&30GHz&44GHz&70GHz\\
 Thickness [mm] & 0.1 & 0.1 & 0.1 \\
 straight part length [mm] & 0.127  & 0.127  & 0.05 \\
 bend  radius [mm] & 0.127   & 0.127  & 0.05 \\
 bend angle [deg] & 0.1  & 0.1  & 0.1 \\ 
 section dimensions [$\mu$m] & 38.1 &25.4 &12.7 \\
    \hline\hline
       \end{tabular}
       \caption{\emph{Mechanical tolerances defined by the manufacturer for the LFI's Copper WGs}}\label{Tab:2}
\end{table}

\begin{table}[hb!]
       \centering
       \begin{tabular}{lcccc}
       \hline\hline
&30GHz&44GHz&70GHz\\
Gold thickness [$\mu$m] & 0.25 & 0.25 & 0.25 \\
SS total length [mm] & 0.127   & 0.127  & 0.127 \\
Gold plated length [mm] & 6.35   & 6.35 & 6.35 \\ 
    \hline\hline
       \end{tabular}
       \caption{\emph{Mechanical tolerances defined by the manufacturer for the LFI's Stainless Steel WGs}}\label{Tab:3}
\end{table}

Even using consolidated manufacturing techniques it was not possible to make the
 Cu section in a single 
piece at higher frequencies. In fact, given the small dimensions and the 
complex routing of the WGs, etching the mandrel was 
extremely slow, about 50 mm/day. Thus, to optimize the formation process while 
preserving the quality of final product,  three 
separate pieces were made at 70 GHz and joined together using the copper joint technique. 
Finally the WGs were washed repeatedly with isopropyl alcohol in order to remove any 
residuals inside them. 
The importance of this last 
step was soon discovered, since in a few occasions a single wash was not enough to 
remove all residuals, but this was apparent only after the electromagnetic 
tests. 

\section{EM testing}
 
The LFI WGs must satisfy many requirements of 
different nature. Despite their complex geometrical shape, they must be 
able to withstand the mechanical stress of launch and for this reason they have been 
submitted to vibration tests. Moreover, their main scope is the 
thermal decoupling of the  FEM and BEM without excessive attenuation of 
the radiometric signal; thus the scattering parameters of each WG and the 
isolation within each set of 4 channels have been measured. The main 
requirements for the electromagnetic properties of the WGs (Guzzi and Villa \cite{guzzi_villa}) are reported in 
table  \ref{Tab:4}.
 \begin{table}[hb!]
       \centering
       \begin{tabular}{lcccccccccccccccccccccccccccc}
       \hline\hline
 &$\nu$ band[GHz]& &IL@20K&&RL&&Isolation\\
 
 30GHz   &   27-33 &&  $<$ 2.5dB &&  $>$ 25dB & & $>$ 30dB   \\
 44GHz    & 39.6-48.4  && $<$ 3dB & &$>$ 25dB & &$>$ 30dB  \\
 70GHz    & 63-77  && $<$ 5dB &&$>$ 25dB && $>$ 30dB   \\
       \hline\hline
       \end{tabular}
       \caption{\emph{Electromagnetic requirements for the LFI's WG at 
operational temperature}}\label{Tab:4}
\end{table}

\begin{figure}[!h]
\begin{center}
\includegraphics[width=0.6\textwidth]{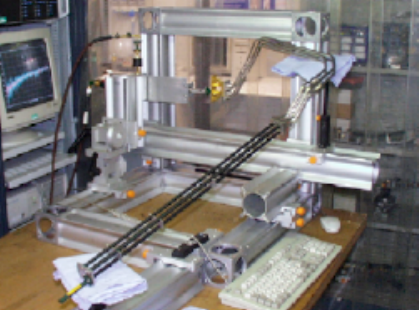}
\caption{\emph{Experimental set up for electromagnetic tests.}} 
\label{fig:1port}
\end{center}
\end{figure}

The detailed electromagnetic characterization of the LFI's WGs 
was a challenging task. 
One of the first constraints is related to the length of the WGs; 
ideally the best solution for measuring the scattering 
parameters would be working in a two port reflectometer configuration. 
But when measuring very large components such as the LFI's WGs, the path 
difference between the calibration standards and the Device Under Test (DUT) is 
very large, and this requires very large cable movements. Since the
measurements were performed using an AB-Millimetre Vector Network 
Analyzer (VNA) which works without an external reference channel, excessive cable movements 
degrade the quality of measurements. For this reason, all tests 
performed on the WGs were made in a one port reflectometer configuration (figure \ref{fig:1port}). 

Another undesired effect of the large path difference between 
DUT and calibration standards, is the 
measurement error arising when the frequency stability of the master 
oscillator is inadequate (or if the VNA operates in swept frequency mode).
A way to avoid this problem was found performing measurements 
with different cable lengths and then combining the results of two 
different measurements (Simonetto et al. \cite{simonetto}). 
Once these main difficulties were solved, the measurement technique was standard. All tests were performed using the 
variable short, variable load and fixed short calibration. After 
calibration, the WGs were terminated with a matched load to determine the 
reflection coefficient and with a short to evaluate the transmission coefficient. 

Working in a one port configuration, the quality of data can be degraded
if return loss and insertion loss are comparable. 
In fact, terminating 
a DUT with an ideal short in a one-port configuration, the measured reflected signal  \emph{b} at port 1 is 

 \begin{equation}
b=S_{11}-\frac{S_{12}\cdot S_{21}}{1+S_{22}}.
\end{equation}

\begin{figure}[!h]
\begin{center}
\includegraphics[width=0.6\textwidth]{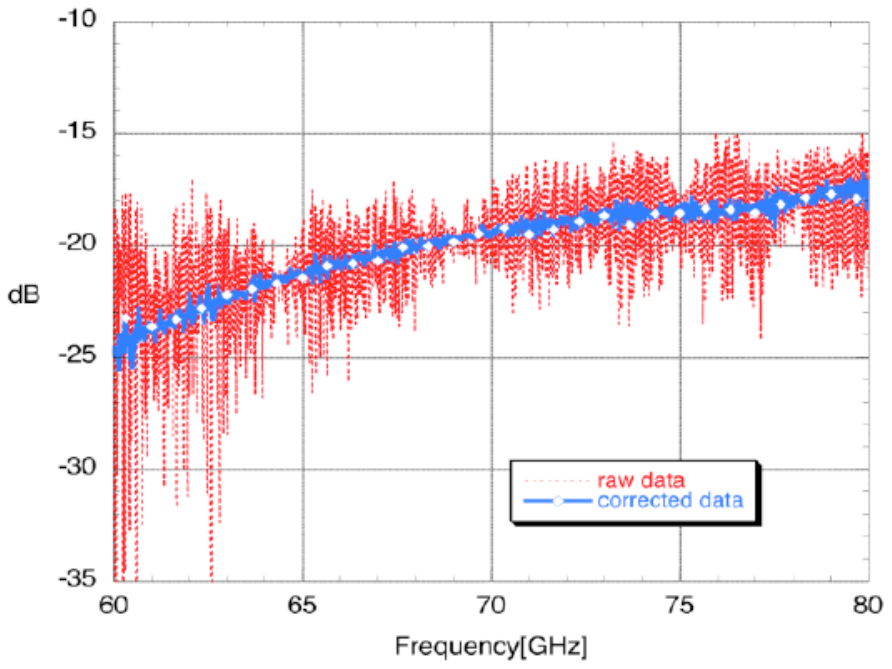}
\caption{\emph{70 GHz transmission coefficient, raw data (dashed line) and corrected data (thick 
solid line with symbols.)}} \label{fig:IL_RLcorr}
\end{center}
\end{figure}

The measured data are therefore a combination of IL and RL, and of course 
the contribution of RL cannot be neglected unless the DUT is low loss and 
very well matched.
Since the return loss was measured, those data were used to correct the measurements by direct substitution, as described in (D'Arcangelo et al. \cite{darcangelo}). This practice is far from optimum in terms of error propagation, therefore the results were compared with those obtained with the more accurate (and time consuming) 
approach of taking the ratio of coefficients in the error matrices extracted from full calibrations with and without DUT.
The two methods are consistent within the error bars,
A comparison between raw and corrected data is shown in fig. \ref{fig:IL_RLcorr}.
The FEM flanges at all frequencies and the BEM 
flanges at 44 and 70 GHz required adapters. In order 
to evaluate losses and reflections of the WGs, it is necessary 
to evaluate the same quantities for the adapters too, thus they were 
tested with the calibration technique described previously. Since 
the losses of the adapters were extremely low and comparable with the 
measurement error (about 0.1 dB), a very precise determination 
of this value within the bandwidth was not possible, but anyway results 
can be considered as an upper limit. 
Adapters have an impact also on the measured return loss of the 
WG: a way to remove or minimize this unwanted contribution is using time
domain digital filtering, known as time gating in analog terms. The Fourier transform
 (FT) of the frequency sweep allows in fact a quick identification of the 
position of the main reflections and their removal if 
not caused by the WG itself. The ideal condition for this 
operation would be a very good space resolution, in order to determine 
with high precision the exact position of the reflection points. The
limiting factors are measurement  bandwidth and waveguide dispersion. However, it was possible to guarantee a satisfying resolution (about 5-8 mm depending on the LFI frequency channel). 
Quite soon, already during the qualification model test campaign, it became
apparent that testing the two different WG sections independently would be highly beneficial.
On the one hand, this allows a detailed 
knowledge of the behaviour of each section of the WG, and on 
the other it helps in determining the location of faults. Thus, after being tested as single units, the Cu and SS WG are integrated and tested again as a single piece in order to guarantee that a good match between the two sections was obtained
Electromagnetic tests turned out to be a very 
useful tool for spotting problems in the WGs. 
Typically, the problems were residuals of the aluminum mandrel that 
was not completely dissolved, and in one case a chemical deposit 
left inside the SS section after Au deposition. 
\subsection{Transmission Coefficient}
\begin{figure}[!h]
\begin{center}
\includegraphics[width=0.6\textwidth]{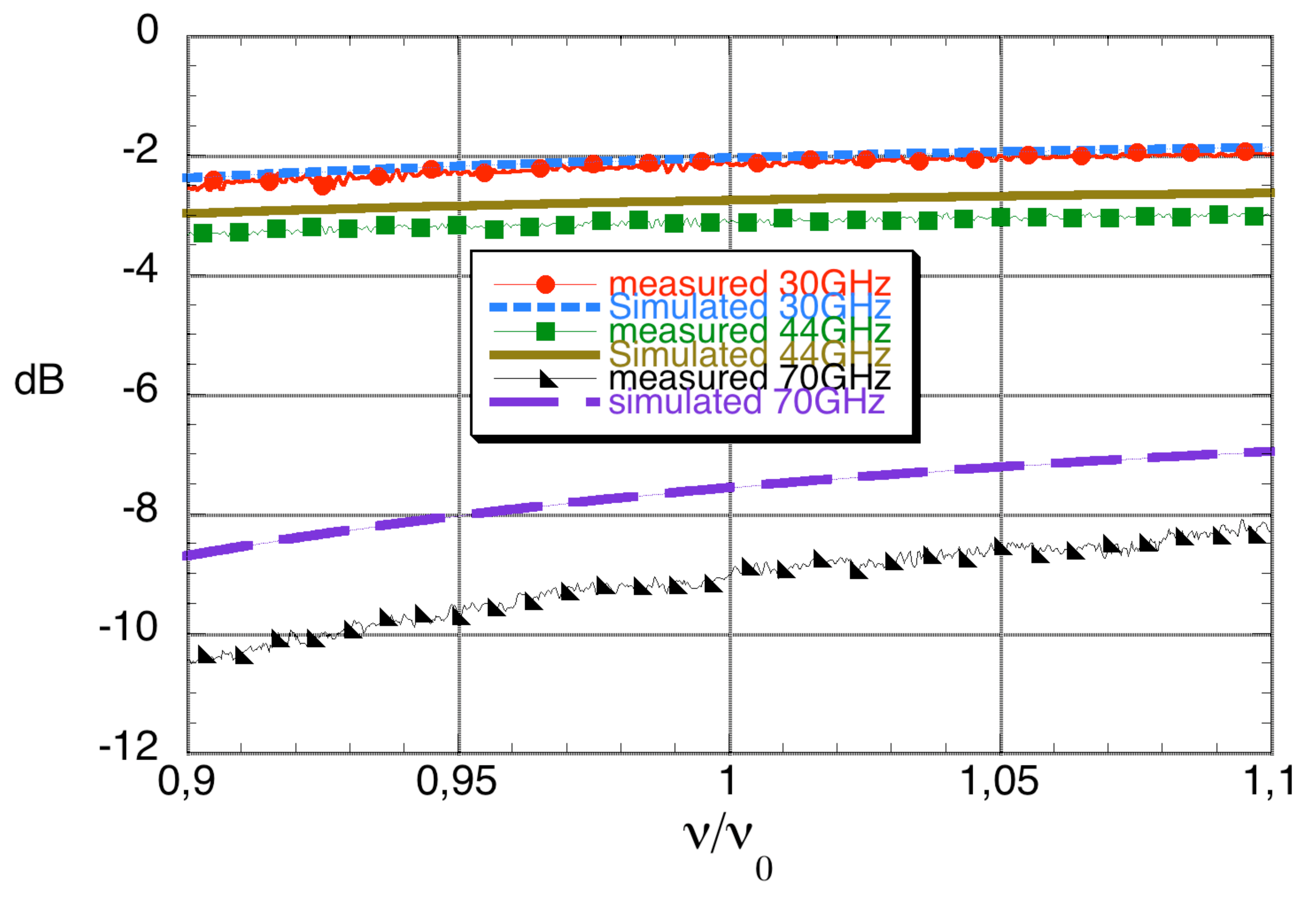}
\caption{\emph{Average amplitude of the transmission coefficient of 4 LFI's WG, compared with the simulated losses 
of a straight rectangular guide.}} \label{fig:IL}
\end{center}
\end{figure}

\begin{figure}[!h]
\begin{center}
\includegraphics[width=0.6\textwidth]{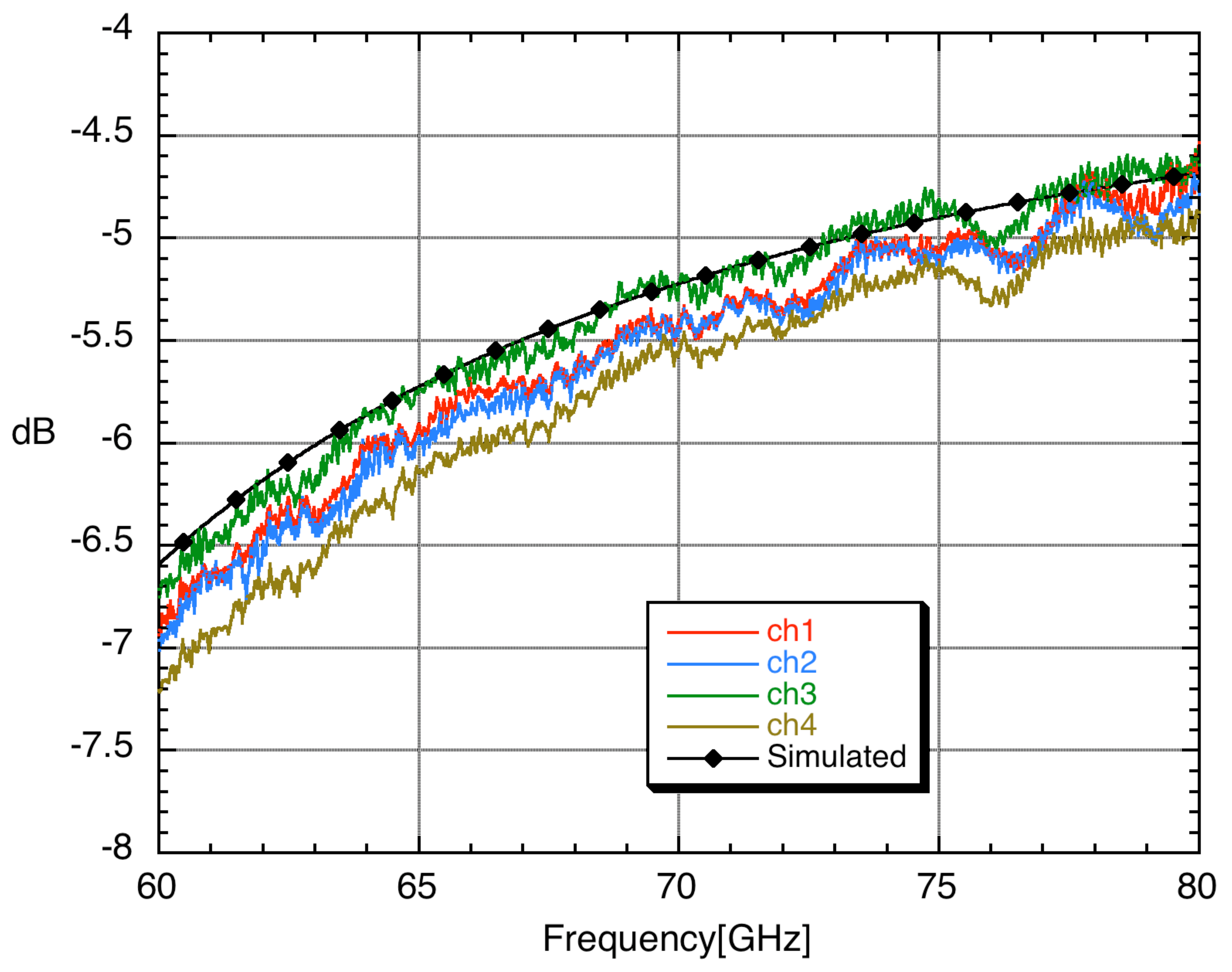}
\caption{\emph{Amplitude of the transmission coefficient of 4 70GHz LFI's SS WGs, compared with the simulated 
losses of a straight rectangular guide.}} \label{fig:IL70SS}
\end{center}
\end{figure}

\begin{figure}[!h]
\begin{center}
\includegraphics[width=0.6\textwidth]{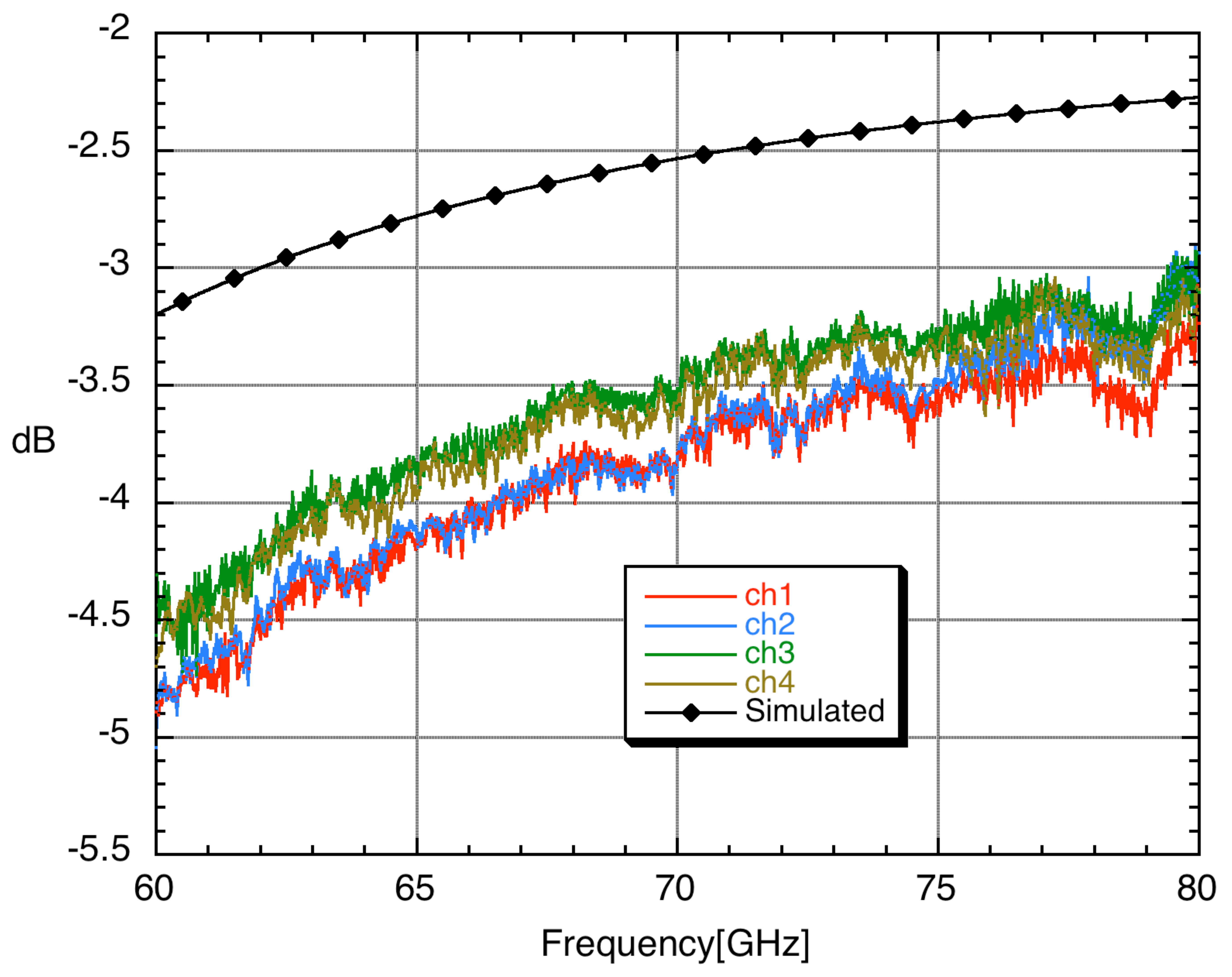}
\caption{\emph{Amplitude of the transmission coefficient of 4 70GHz LFI's Cu WGs, compared with the simulated 
losses of a straight rectangular guide.}} \label{fig:IL70Cu}
\end{center}
\end{figure}

The transmission coefficient of the WGs was determined by subtracting the contribution 
of the flange adapters, even though it was usually much smaller than the WG loss. 
At 30 GHz only one adapter was used, since the BEM-side flange was compatible with the standard UG 599 ones. 
The average amplitude of the transmission coefficient of the 4 WGs belonging to the same bunch is shown in fig. \ref{fig:IL} for the Cu and SS sections in the 3 LFI frequency bands. The frequency is normalized to the central one.
In this plot, data are shown only on the nominal frequency band, but  measurements were taken for all WG over the 
largest possible frequency band, typically 26.5-40 GHz, 33-50 GHz and 
60-80 GHz.
These results are quite representative of the 
general behaviour of all the LFI's WGs. Usually, the measured losses of 
WGs are quite similar (differences of 
fractions of dB) within the same bunch, the difference being mainly due to the different length of the Cu section (a few 
centimeters). 
No requirements having been defined for the amplitude of the transmission coefficient at room temperature, a partial indication of the correct behaviour is given by the comparison with simulated losses.
The result can only be considered partial, since simulations were made considering a standard 
straight rectangular WG, made of gold for 400 mm, SS for 300 mm, and Cu for the rest of its 
length. Bends and twists were considered as a straight sections with equivalent length with the 
pure $TE_{10}$ mode propagating inside. Ideal conductivity was used and the roughness of the internal
walls was neglected. The data usually showed that the difference between simulations and 
measurements is mainly caused by the Cu sections. Typical behaviour is reported in figures
 \ref{fig:IL70SS} and \ref{fig:IL70Cu},  where the losses of the two sections of a 70GHz WG (where the difference is maximum) are compared with simulations. 
Figures \ref{fig:IL70SS} and \ref{fig:IL70Cu} clearly show that the maximum difference between simulations and measurements is 
smaller than 0.5 dB for the SS section, while the discrepancy is greater for the Cu section, even
more than 1.5 dB. This should not be surprising since simulations were made assuming a perfect 
straight waveguide with ideal conductivity. While return loss simulation techniques of lossless
waveguide discontinuities are well established, dedicated simulations to evaluate the insertion 
loss were not addressed because of the intrinsic difficulty to simulate long (in terms of 
wavelenght) dissipative roughness structures. Moreover the simulation would provide an approximate results anyway, since the ohmic losses depend critically on the quality of the inner surface that could only be evaluated with prototyping.
It was then decided to evaluate the effective attenuation coefficient directly from measurments. The analytical simulation where used to 
demonstrated that mainly the $TE_{10}$ mode propagates inside the copper section since the data
and the simulations show the same behaviour. 
While the pictures at 30 
and 44 GHz contain basically the raw data (only the adapter's
contribution having been removed from data), at 70GHz the data were also 
corrected for the contribution of reflections, as described above, since they are not negligible in this case (figure \ref{fig:IL_RLcorr})
Even if specifications were given only at operational temperature, a few considerations can be contemplated. All 30 GHz WGs meet requirements already at room temperature,  and losses are expected to decrease at operational temperature. At 44 GHz, a few WGs meet (at room temperature) the insertion loss requirement of 3 dB (at cryogenic temperature), on a limited portion of the bandwidth, while a few others are very near this value. At 70 GHz, instead, results are far from the requirement value of 5 dB. A rough extrapolation of losses to operational temperature was made considering that, for a straight rectangular WG, the attenuation A, measured in dB, is
 \begin{equation}
A_{dB}=-20\cdot\alpha \cdot l\cdot \log(e)
\end{equation}
where \emph{l} is the WG's length while the attenuation coefficient $\alpha$ is given by
 \begin{equation}
\alpha=\frac{\frac{2}{a}\left(\frac{\pi\epsilon_{0}\nu}{\sigma}\right)^{1/2}}{\left(1-\frac{c^{2}}{4a^{2}\nu^{2}}\right)^{1/2}}\cdot \left(1+\frac{c^{2}}{4a^{2}\nu^{2}}\right)
\end{equation}

where $\nu$ is the frequency, $\sigma$=1/$\rho$ is the conductivity of the material, $\epsilon_{0}$ is the vacuum dielectric constant, $a$ is the broad side of the WG and $c$ is the speed of  light in vacuum.

Data at room temperature were fitted using this formula, thus finding an effective average resistivity of the material at room temperature. This value is compared with the theoretical one to determine the deviation from ideal case. Performance at operational temperature is then obtained considering the WG as if it were composed of small sections at different constant temperature and assuming the resistivity obtained from the fit to scale as function of temperature like the ideal value. 

\begin{figure}[!h]
\begin{center}
\includegraphics[width=0.6\textwidth]{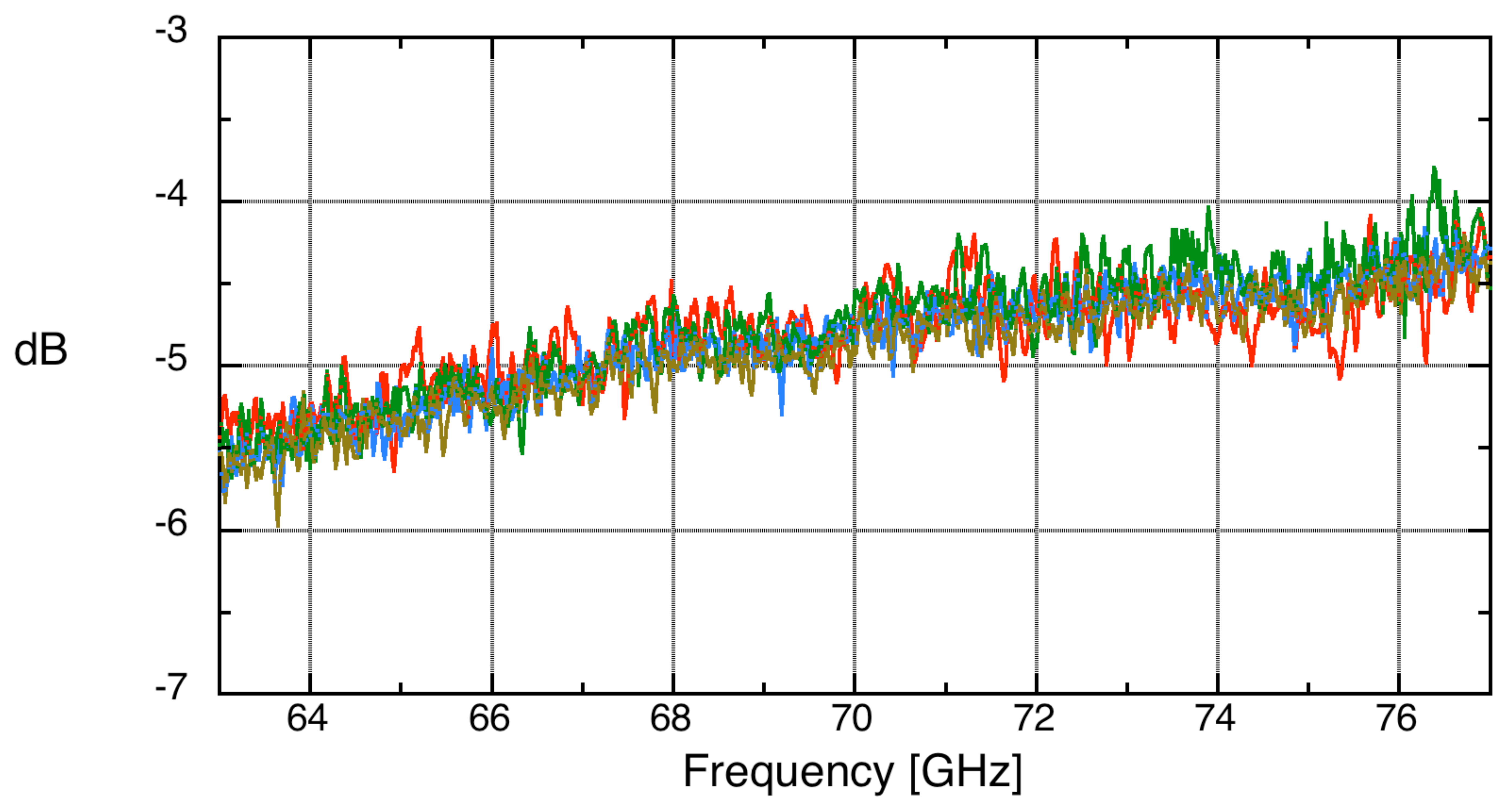}
\caption{\emph{Transmission coefficient measured at T room, scaled at operational temperature.}} 
\label{fig:IL_Tcryo}
\end{center}
\end{figure}

Results are shown in figure \ref{fig:IL_Tcryo}, and they are representative of the 
general behaviour found for the 70 GHz WGs. Extrapolated losses thus match the requirements in this band too. 

\begin{figure}[!h]
\begin{center}
\includegraphics[width=0.6\textwidth]{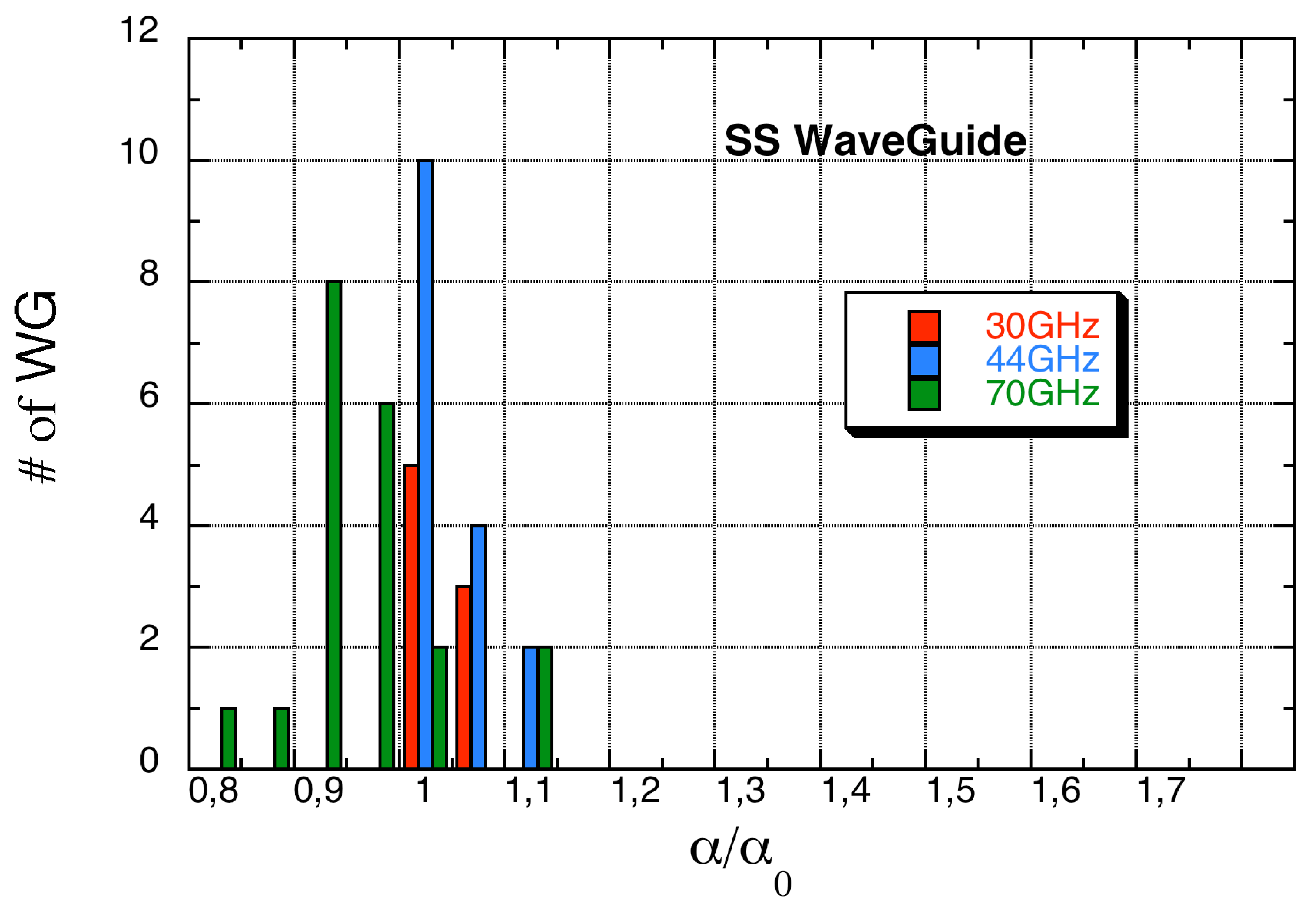}
\caption{\emph{Measured-to-simulated attenuation coefficient ratio of the Stainless Steel WGs}} \label{fig:istogram_SS_WG}
\end{center}
\end{figure}

\begin{figure}[!h]
\begin{center}
\includegraphics[width=0.6\textwidth]{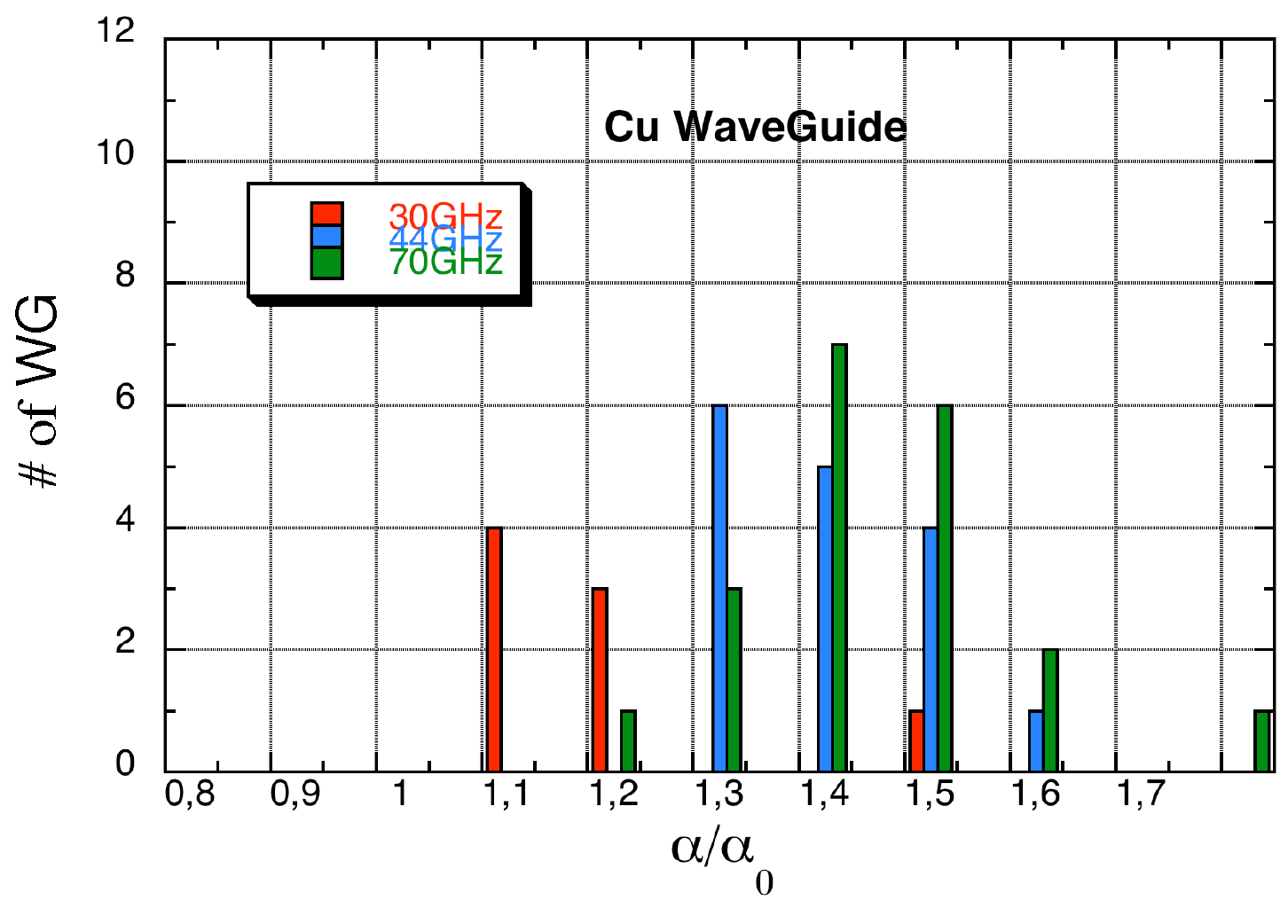}
\caption{\emph{Measured-to-simulated attenuation coefficient ratio of the Copper WGs}} \label{fig:istogram_Cu_WG}
\end{center}
\end{figure}

\begin{figure}[!h]
\begin{center}
\includegraphics[width=0.6\textwidth]{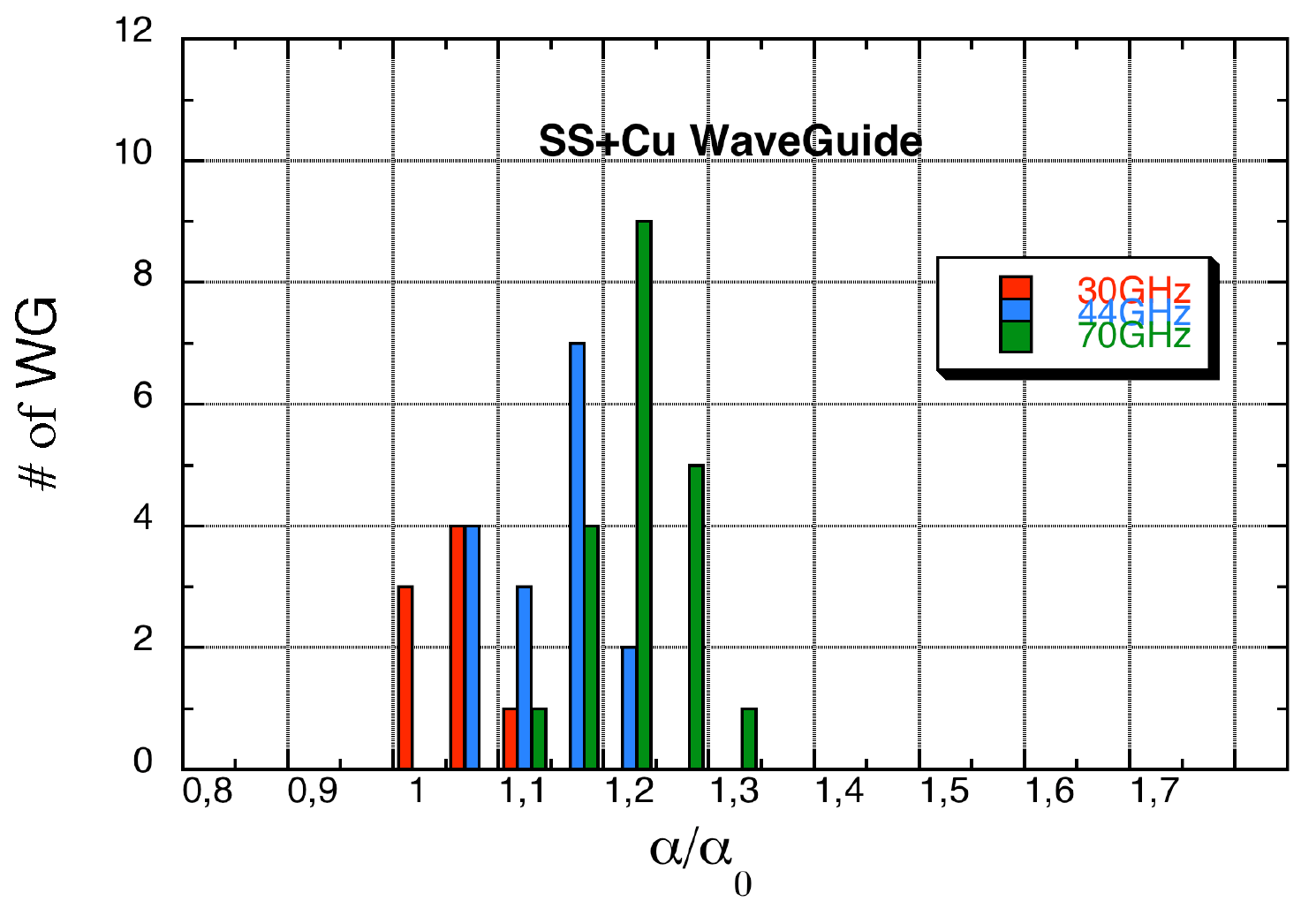}
\caption{\emph{Measured-to-simulated attenuation coefficient ratio of the integrated Stainless Steel and Copper WGs}} \label{fig:istogram_SSCu_WG}
\end{center}
\end{figure}

A statistical analysis was also performed on all the LFI WGs, making a histogram of the ratio between measured and simulated attenuation coefficient. 
The mean value of the frequency-dependent attenuation coefficient was considered for the analysis. The results point out the general behaviour of the WGs, which seems very similar within each band, as shown in pictures \ref{fig:istogram_SS_WG}, \ref{fig:istogram_Cu_WG}, \ref{fig:istogram_SSCu_WG}. 

The SS WGs have attenuation coefficients close to the simulated one, and this is expected since they are all straight rectangular waveguides. On the other hand for the Cu WGs the value $\alpha$/$\alpha_{0}$ is centered around 1.4. Of course, simulations do not take into account twists and bends, but even an incorrect assumption of the conductivity of the material (assumed ideal, with no account for surface roughness) could explain the difference. Also, the spread of the results for Cu is broader than for SS section and this could point to a less reproducible manufacturing process, but the impact of twists and bends should be assessed thoroughly before drawing such a conclusion. 
The integrated WGs have of course an intermediate behaviour and this is an indication that the integration process does not introduce anomalies. The distributions of figures \ref{fig:istogram_SS_WG}, \ref{fig:istogram_Cu_WG}, \ref{fig:istogram_SSCu_WG} were fitted with a gaussian function: the results of the fit are listed in table \ref{Tab:5}.

 \begin{table}[hb!]
       \centering
       \begin{tabular}{lccc}
       \hline\hline
WG &$\alpha/\alpha_0$&$\sigma$\\
 Stainless Steel section &  0.99  & 0.064   \\
 Copper Section &1.4 & 0.13  \\
 Integrated WG& 1.15& 0.086   \\
       \hline\hline
       \end{tabular}
       \caption{\emph{Results of the gaussian fit for the attenuation coefficient.}}\label{Tab:5}
\end{table}

\subsection{Reflection coefficient}
\begin{figure}[!h]
\begin{center}
\includegraphics[width=0.55\textwidth]{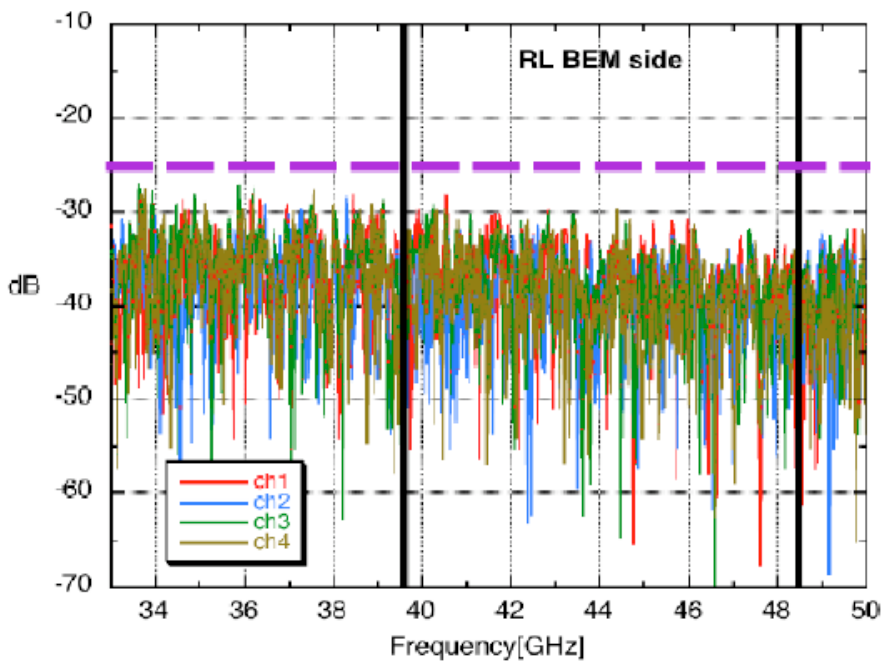}
\caption{\emph{Amplitude of the reflection coefficient at 44 GHz, signal inserted from the BEM flange. The two 
vertical lines represent LFI's frequency band, while the horizontal 
one represents the room T requirement.}} \label{fig:RL_Bem44}
\end{center}
\end{figure}

\begin{figure}[!h]
\begin{center}
\includegraphics[width=0.55\textwidth]{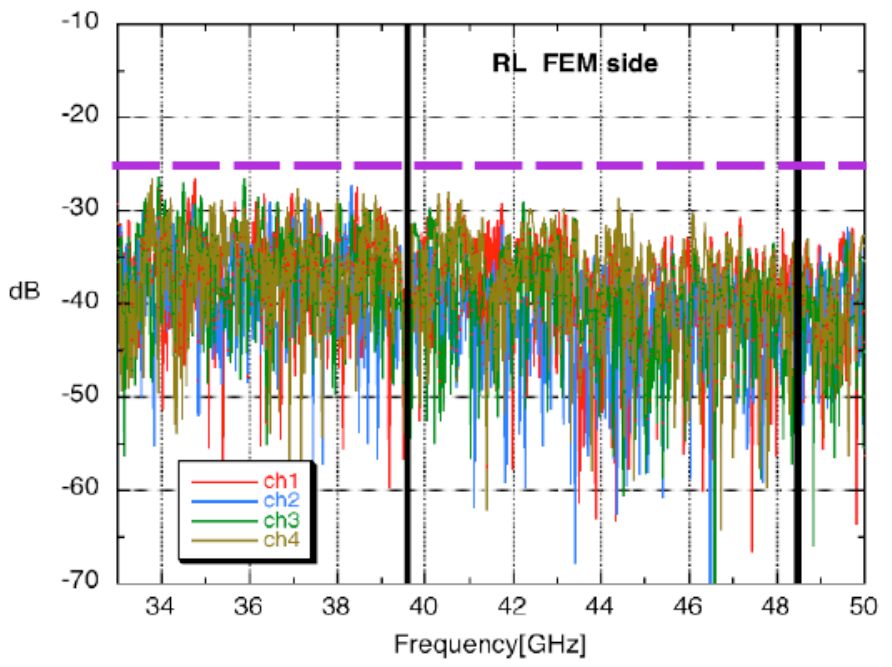}
\caption{\emph{Amplitude of the reflection coefficient at 44 GHz, signal inserted from the FEM flange. The two 
vertical lines represent LFI's frequency band, while the horizontal 
one represents the room T requirement.}} \label{fig:RL_Fem44}
\end{center}
\end{figure}

\begin{figure}[!h]
\begin{center}
\includegraphics[width=0.55\textwidth]{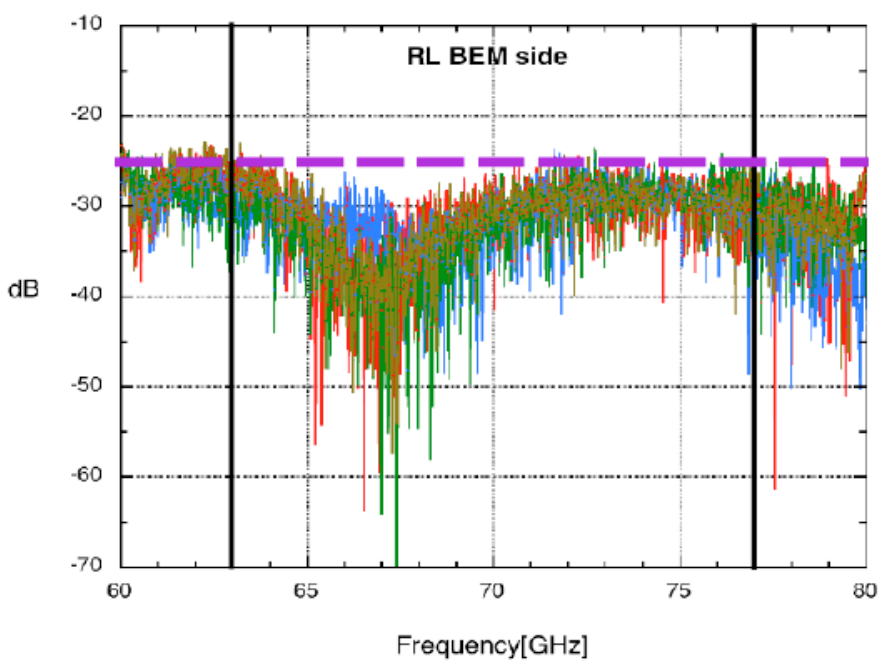}
\caption{\emph{Amplitude of the reflection coefficient at 70 GHz, signal inserted from the BEM flange. The two 
vertical lines represent LFI's frequency band, while the horizontal 
one represents the room T requirement.}} \label{fig:RL_Bem70}
\end{center}
\end{figure}

\begin{figure}[!h]
\begin{center}
\includegraphics[width=0.55\textwidth]{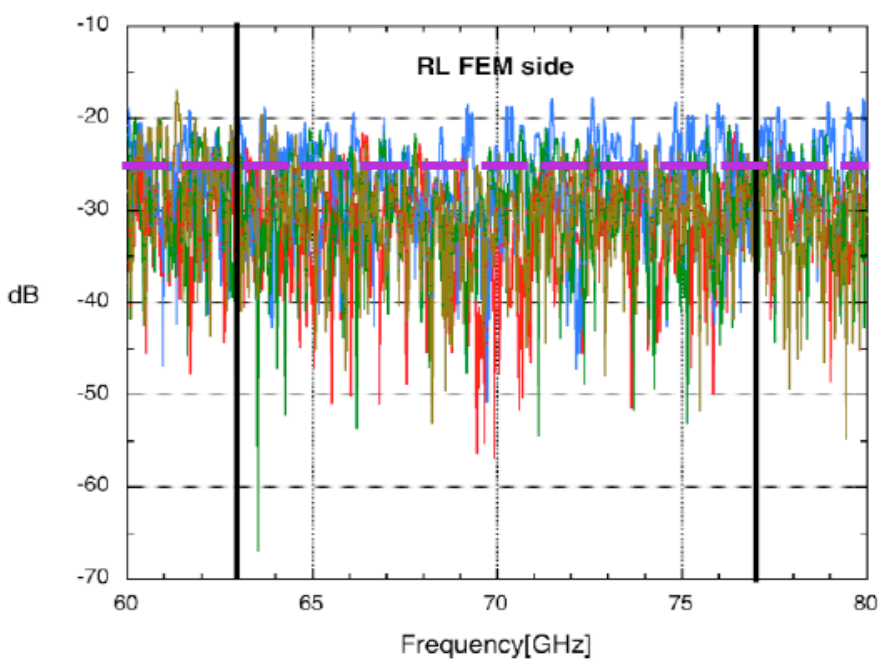}
\caption{\emph{Amplitude of the reflection coefficient at 70 GHz, signal inserted from the FEM flange. The two 
vertical lines represent LFI's frequency band, while the horizontal 
one represents the room T requirement}} \label{fig:RL_Fem70}
\end{center}
\end{figure}

Representative results for the 70 GHz and the 44 GHz frequency 
bands are shown in figures  \ref{fig:RL_Bem44}, \ref{fig:RL_Fem44}, \ref{fig:RL_Bem70} and \ref{fig:RL_Fem70} (results at 30 GHz are extremely similar to those at 44GHz). 
%\newpage
Since losses at 30 and 44 GHz are similar and not too high, the return loss measured at the FEM and BEM side is similar; in fact the signal reflected at any discontinuity inside the waveguide is not attenuated significantly on the way back to the injection port.  At 70 GHz instead, where WG losses are definitely greater,  the difference between the FEM and BEM side is clearly visible. In fact, the main sources of 
reflection are generally in the Cu section. In particular, the regions with the narrowest bends are the most reflective ones. This can be seen when correcting 
reflection data using time domain filtering (time--gating): the FT of a frequency sweep gives a sequence of peaks as a function of time, representing reflections from
different parts of the device under test. This clearly points out that main reflection peaks are located in the CU section of the WGs, as expected. The FT is also use to filter (i.e. zeroing out) the unwanted contribution of adapters and transforming back to frequency allows correction of the data.
Results show that requirements are met with some margin at 30 and 44 GHz, while at 70GHz the amplitude of the reflection coefficient is around the requirement limit of -25dB for all WG, especially at the FEM side.
%\newpage
%\newpage

\subsection{Isolation}
\begin{figure}[!h]
\begin{center}
\includegraphics[width=0.55\textwidth]{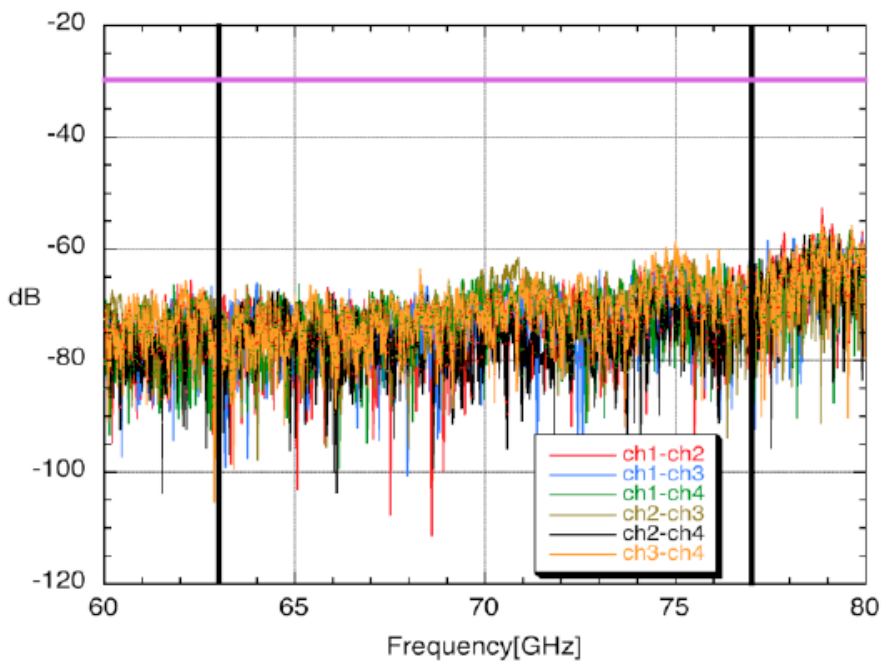}
\caption{\emph{70 GHz isolation, the two vertical lines represent the 
LFI's frequency band, while the horizontal one represents the T room 
requirement.}} \label{fig:iso70}
\end{center}
\end{figure}

The last step in the electromagnetic characterization of the LFI's WGs is 
the evaluation of isolation between the 4 WGs belonging to the same bunch. The measurement is made in a two ports 
configuration with only a simple response correction calibration, basically because the expected measured level is very low 
and an overly accurate measurement is not necessary. 
The requirement 
at room temperature is -30 dB, and results obtained in the 70 GHz band are
reported in figure \ref{fig:iso70}. They are representative also for the other bands. 
That figure clearly points out that the measured 
isolation is very high,  dominated by instrumental noise and well below the requirement.
This is true for all LFI's flight WGs, and is expected to hold true also at cryogenic temperature.
\newpage
\section{Vibration tests}
 
The LFI WGs must be vibrated at space qualification level but, given 
their shape and physical dimensions, it was not possible to vibrate the Cu and SS 
section while integrated. The two sections have been  tested separately: each WG
has two flanges which, during operations, are submitted to two different dynamic environments. Since the experimental set up did not allow simultaneous excitation of random vibrations at two different locations, each WG was vibrated with an input random level (along the three axis) corresponding to the envelope of those relevant to the two flanges. 

\begin{figure}[!h]
\begin{center}
\includegraphics[width=0.55\textwidth]{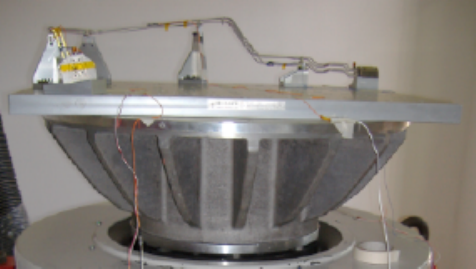}
\caption{\emph{70 GHz Cu WG and fixture mounted over the shaker}} 
\label{fig:Cu70vibrfix}
\end{center}
\end{figure}

Vibration tests were performed with a shaker (figure \ref{fig:Cu70vibrfix}) controlled by a computer, which generates the motion with the requested characteristic.The  accelerometers, placed along the WG and its flanges, were used for feedback. The accelerometers' recorded signals were amplified, digitized and stored for processing.

\begin{figure}[!h]
\begin{center}
\includegraphics[width=0.65\textwidth]{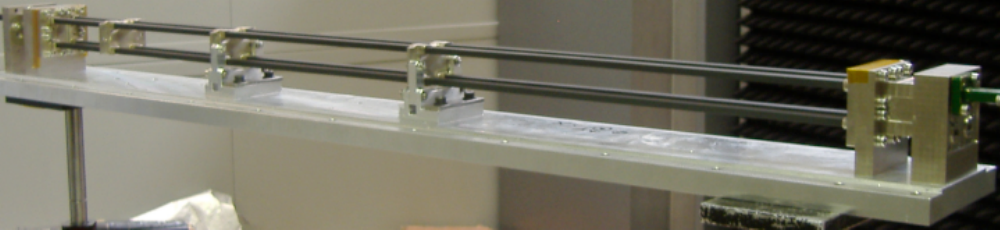}
\caption{\emph{30 GHz SS WG mounted over the vibration fixture.}} 
\label{fig:SSvibr}
\end{center}
\end{figure}

\begin{figure}[!h]
\begin{center}
\includegraphics[width=0.65\textwidth]{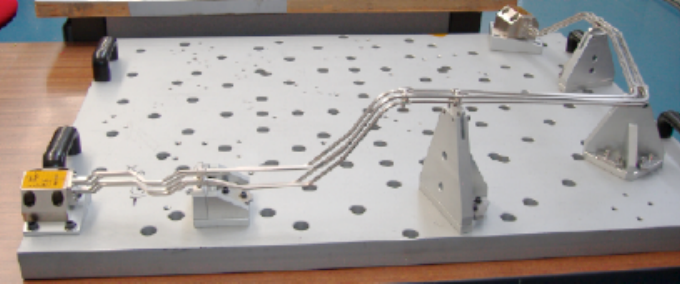}
\caption{\emph{70 GHz Cu WG mounted over the vibration fixture.}} 
\label{fig:Cuvibr}
\end{center}
\end{figure}

It was necessary to manufacture two different fixtures (one for the SS WGs, figure \ref{fig:SSvibr}, and one for the Cu WGs, weighting 50Kg, figure \ref{fig:Cuvibr}) and some specific flange supports for this test. The supports have a double use. On the one hand they were used for solid connection with the vibrating structure, and on the other hand they allow checking the resiliency of flange joints to vibration. They consist of aluminum blocks with four RF channels each, thus the two blocks and the WG form a single unit of four RF channels. 

The test started with a search for resonances in the range 5-2000Hz once the WG was firmly mounted on the shaker.
The following conditions had to be satisfied for a successful vibration test: after each random vibration on every axis, the WG should not show visible degradation and all the screws holding it to the fixture should be tight to specified torque. Moreover, the mechanical resonance frequencies before and after vibration should be shifted by less than ±5\%, while the variation in acceleration should be smaller than ±10\%. 

\begin{figure}[!h]
\begin{center}
\includegraphics[width=0.55\textwidth]{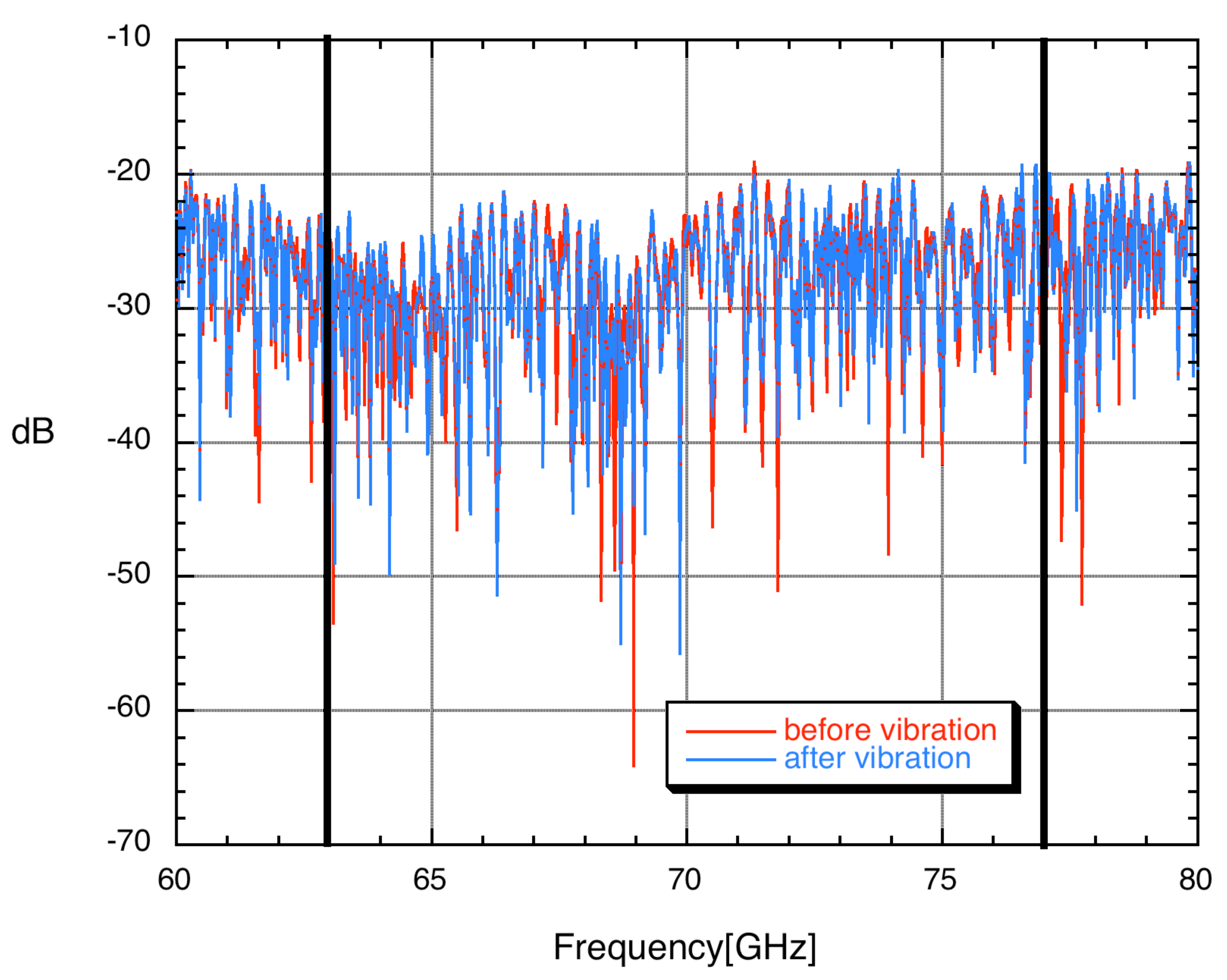}
\caption{\emph{Reflection coefficient of a Cu 70 GHz WG measured before and after the vibration tests}} 
\label{fig:RL_pre_post_vib}
\end{center}
\end{figure}

Finally, electromagnetic tests were repeated to certify success in vibration test. The reflection coefficient of the WG was controlled before 
and after vibration, because of its sensitivity to mechanical 
imperfections. These measurements were repeated inserting the signal from both sides, to guarantee that the electromagnetic effect of any damage was not hidden by the attenuation of the WG. In the very few cases where differences were seen, they were found in the coupling between the aluminium blocks and the adapters use to connect to the microwave circuit. An example of the comparison between the reflection coefficient measured before and after vibrations is reported in figure \ref{fig:RL_pre_post_vib}.
Moreover also the Fourier Transform of the 
signal measured before and after the vibration was calculated and 
compared. In this way any possible variation in the reflections peaks 
caused by the vibration could be detected. 

12 SS and 12 Cu WGs (i.e. 1 bunch of 4 WGs for each frequency) have been 
vibrated: the very first tests, performed at  low energy, suggested a structural weakness of the WGs. Thus, all the flanges were reinforced, and all the WGs were 
retested afterwards, to guarantee that reworking did not 
impact on performance. After these steps, the 12 SS and 12 Cu WGs 
were successfully vibrated. Only Flight Spare (FS) components were vibrated, in order to avoid excessive stress on the flight components. 
 
 \section{Summary}
Six prototypes, 5 Qualification Model, 22 FM and 22 FS WGs were built and extensively tested. They include a straight SS section, partially gold plated, and a Cu section with twists and bends along. Because of 
their peculiar geometrical shape and physical dimensions, it was 
necessary to optimize the measurements technique in order to obtain a 
precise characterization. The manufacturing technique was also
tested and optimized during the qualification phase. 
Electromagnetic requirements are usually met. A mismatch between 
simulated and measured losses at room temperature was observed, increasing with frequency. Data 
collected over the two separate sections clearly pointed out that the 
difference is basically due to the Cu part, and it should be accounted on the oversimplified simulations neglecting twists and bends. 
Finally, all FM WGs were successfully vibrated to space 
qualification level, although flanges needed reinforcement.  

\begin{center}
 \textbf{acknowledgements}
\end{center}
Planck is a project of the European Space Agency with instruments
funded by ESA member states, and with special contributions from Denmark
and NASA (USA). The Planck-LFI project is developed by an International
Consortium led by Italy and involving Canada, Finland, Germany, Norway,
Spain, Switzerland, UK, USA. The Italian contribution to Planck is
supported by the Italian Space Agency (ASI).

\end{document}